\documentclass[12pt,color]{article}
\usepackage{amscd}
\usepackage{amsmath}
\usepackage{amsfonts}
\usepackage{amssymb}
\usepackage{color}
\usepackage{epsfig}
\usepackage{subeqnarray}
\usepackage{afterpage}

\allowdisplaybreaks[4]

\begin{document}

\title{Scaling properties of the pairing problem in the strong coupling limit}

\author{M. B. Barbaro$^1$, R. Cenni$^2$,
A. Molinari$^1$
and M. R. Quaglia$^2$\\
${}^1$ Dipartimento di Fisica Teorica --- Universit\`a di Torino
\\
Istituto Nazionale di Fisica Nucleare --- Sez. di Torino\\
 Torino ---Italy \\
${}^2$ Dipartimento di Fisica --- Universit\`a di Genova\\
Istituto Nazionale di Fisica Nucleare --- Sez. di Genova\\
Genova --- Italy\\}

\date{}

\maketitle

\begin{abstract}
  We study the excited states of the pairing Hamiltonian 
  providing an expansion for their energy in the
  strong coupling limit. 
  To assess the role of the pairing interaction
  we apply the formalism to the case of a heavy atomic nucleus.
  We show that only a few statistical moments
  of the level distribution are sufficient to yield an accurate estimate
  of the energy for not too small values of the coupling $G$
  and we give the analytic expressions of the first four terms of the series.
  Further,  we discuss the convergence radius $G_{\rm sing}$ of the
  expansion showing that it strongly depends upon the details of the level 
  distribution. Furthermore $G_{\rm sing}$ is not related to the critical
  values of the coupling $G_{\rm crit}$, which characterize the physics of
  the pairing Hamiltonian, since it can exist even in the absence of these
  critical points.
\end{abstract}

\section{Introduction}

The problem of the pairing interaction in a Fermi system, like e.g. an atomic 
nucleus, has been dealt with long time ago in the case of $n$ pairs living in 
a single level and the solution is well-known.

Instead, the case of $n$ pairs distributed over a set of $L$ levels,
each one with a pair degeneracy $\Omega_\mu$ and energy $\epsilon_\mu$,
is treated with the Richardson's equations \cite{Ri-65}
(in short RE), to be described below, but no explicit solution can be given 
in a closed form: hence 
for a finite system one has to resort to numerical methods \cite{Rombouts}.

The RE, assuming that $n$ pairs are distributed, in the absence of
interaction, over $L$ levels, read
\begin{equation}
  \label{eq:A001}
  \sum_{\mu=1}^L\frac{\Omega_\mu}{2\epsilon_\mu-E_i}
  -2\sum_{k\not=i}^n\frac{1}{E_k-E_i}=\frac{1}{G}
\end{equation}
with $i=1,\dots,n$, $G$ being the strength of the interaction. 
The $E_i$, namely the unknowns, are sometimes viewed as the energies of the
single ($i^{th}$) pairs, although
this statement is questionable: in fact they have no direct
physical meaning and could be complex. The true observable, namely the
energy of the system in a given state, in the Richardson framework turns out
to be
\begin{equation}
  \label{eq:A002}
  E=\sum_{i=1}^n E_i
\end{equation}
and is of course real. In \eqref{eq:A001} the $n$ pairs are set up by two
fermions in time reversal states coupled to zero momentum or angular momentum.

The space of the parameters in the pairing problem is wide, being generated
by the set of the unperturbed s.p.e. (single particle energies)
$\{\epsilon_\mu\}$, by their pair degeneracies $\{\Omega_\mu\}$ and by the
coupling constant $G$. We remind that $L$ can be of the order of, say,
10 or less in the nuclear case, but is of the order of the Avogadro number
in the case of a band in a metal.

In the strong coupling limit, however, the dependence upon
the whole set of parameters occurs  only through some simple 
combination of them.
Actually here the RE reduce to a system of equations essentially 
parameter-free,
whose solutions, namely the $E_i$, can be obtained by exploiting a scaling 
property, hence an analytic solution for the total energy can be given,
although not in a closed form, but as an expansion in inverse powers of $G$.

This topic has been addressed in some recent 
papers~\cite{Yuz,Snyman,CC} in the context of solid state
physics (actually superconducting metallic grains) where a major simplification
occurs since the unperturbed levels are assumed to be
equally spaced ($\epsilon_\mu=\hbar\omega_0\mu$
with $\mu=1,2,\cdots L$) and to host only one pair ($\Omega_\mu=1$). 

The case of nuclear physics requires an 
extension of this approach since the nuclear levels are
distributed inside a major shell with various energies and degeneracies.
The pairing problem for $n$ pairs living in any number of degenerate
levels has been recently addressed in the framework
of pseudodeformed quasispin $SU(2)$ algebra~\cite{Baer} and an exact solution 
has been provided in Ref.~\cite{BP} for an orbit-dependent interaction in the 
special case of two nondegenerate energy levels, but an analytic solution
to the general problem is not presently available.

As we shall see, however, 
in the strong coupling limit the pairing energy depends only
upon the statistical properties of the level distribution.

In this paper we propose a further derivation of the strong coupling
expansion which, 
extending the one presented in Ref.~\cite{Barbaro:2006rc}, 
applies not only to the ground state, but to the excited states energies as well 
and neatly displays in its coefficients the renormalization
of the statistical moments of the levels distribution, of the strength of
the interaction and of the number of pairs prevented to take an active part 
into the dynamics induced by the interaction with the trapped pairs. 
Moreover, and importantly, we succeed as in Ref.\cite{RMP} in yielding analytic 
expressions for the ``pair energies'' $E_i$.
This we do in Sections 2 and 3.
In Section 4 and 5 we address a specific nuclear problem to illustrate how the
method works and in Section 6 we compare our results with the exact numerical
solution of the RE.

\section{The strong coupling expansion}

\label{sec:1emezzo}

Let us first define the ``strong coupling limit''.
As already discussed in \cite{BaCeMoQu-02,BaCeMoQu-04} and \cite{Barbaro:2006rc}, we introduce 
the average 
\begin{equation}
  \label{eq:A003}
  \bar\epsilon=\frac{1}{\Omega}\sum_{\mu=1}^L\Omega_\mu\epsilon_\mu
\end{equation}
and the variance 
\begin{equation}
  \label{eq:A004}
  \sigma=\sqrt{\frac{1}{\Omega}\sum_{\mu=1}^L\Omega_\mu
    \left(\epsilon_\mu-\bar\epsilon\right)^2}~,
\end{equation}
with
\begin{equation}
  \label{eq:A005}
  \Omega=\sum_{\mu=1}^L\Omega_\mu~,
\end{equation}
of the levels distribution.
Then, since $\bar\epsilon$ is an intrinsically
irrelevant parameter, as it depends upon the choice of the zero point of the
 energy, we safely assume $\bar\epsilon=0$.
Thus the two energy scales entering into the pairing problem will be
set by $\sigma$ and $G$.
When the condition $\sigma\ll G$ is met, then the single particle levels 
(s.p.l.) span a
very narrow energy range and the well-known solution
\begin{equation}
  \label{eq:A006}
  E=-G\left(n-\frac{v}{2}\right)\left(\Omega-n-\frac{v}{2}+1\right)
\end{equation}
is expected to be a very good approximation ($v$ denotes the seniority).

It was found in \cite{BaCeMoQu-02} that a convenient expansion parameter is
\begin{equation}
  \label{eq:A007}
  \alpha=\frac{2\sigma}{G\Omega}~,
\end{equation}
the strong coupling limit corresponding to
\begin{equation}
  \label{eq:A008}
  \alpha\ll 1~.
\end{equation}

It is known that 
in the strong coupling limit for a given state some of the $E_i$ are large
(those contributing to the collectivity of the state) 
and of the order of $G$, while the others remain
trapped between the unperturbed levels and are consequently of the order
of $2\sigma$.
Actually the collectivity is associated with th existence or not of broken pairs. The state with zero seniority is {\em the collective state} and has the lowest
energy. The other states, with $v=2,4,\cdots$, correspond to larger energies,
but the broken pairs still contribute to the energy through the Pauli principle.

Hence, having chosen $\bar\epsilon=0$ it is natural to split, introducing an 
integer $k$,  the solutions $E_i$ into two subsets 
$\{E_i,i=1,\dots,k\}$ and $\{E_j,j=k+1,\dots,n\}$ with the condition
$|E_i|\ll | E_j|~\forall i,j$. Thus in this partition the first $k$ pairs are
trapped while the others take part in setting up the collective excitations 
of the system.
Clearly $k$ just corresponds to the Gaudin number $N_G$~\cite{Gaudin},
which in turn is 
related to the ``like-seniority'' $v_l$ introduced in \cite{BaCeMoQu-04}
according to $N_G=v_l/2$.

Consider then the equations for the $E_j$ with $j=k+1,\dots,n$ (the ``large'' 
energies):
by expanding in powers of the small quantities $E_i$ and $\epsilon_\mu$
we get
\begin{equation}
  \label{eq:A009}
  -\frac{1}{E_j}\sum_{\mu=1}^L\Omega_\mu\sum_{m=0}^\infty
  \left(\frac{2\epsilon_\mu}{E_j}\right)^m
  +\frac{2}{E_j}\sum_{i=1}^k\sum_{m=0}^\infty
  \left(\frac{E_i}{E_j}\right)^m
  -2\sum_{\substack{p=k+1\\ p\not=j}}^n\frac{1}{E_p-E_j}=\frac{1}{G}~.
\end{equation}

Likewise the first $k$ equations, related to the trapped solutions, 
can be expanded as follows
\begin{equation}
  \label{eq:A011}
  \sum_{\mu=1}^{L}\frac{\Omega_\mu}{2\epsilon_\mu-E_i}-2
  \sum_{\substack{p=1\\ p\not=i}}^k\frac{1}{E_p-E_i}=\frac{1}{G}
  +2\sum_{p=k+1}^n\frac{1}{E_p}\sum_{m=0}^\infty\left(\frac{E_i}{E_p}
    \right)^m~.
\end{equation}
The equations \eqref{eq:A009} and \eqref{eq:A011} are of course exact providing
the above expansions converge. 

We introduce next the statistical
moments of the levels distribution according to~\footnote{Note that the
present definition differs from the one of Ref.~\cite{Barbaro:2006rc}
 by the factor 
$\sigma^n$.}
\begin{equation}
  \label{eq:B010}
  M^{(n)}=\sigma^nm^{(n)}=\frac{1}{\Omega}
  \sum_{\mu=1}^L\Omega_\mu\epsilon_\mu^n
\end{equation}
(note that $m^{(1)}\propto\bar\epsilon=0$,
$m^{(2)}\equiv 1$ by definition and, of course,
$m^{(0)}=1$)
and rewrite the unknown $E_i,E_j$
in terms of the new dimensionless variables $z_i,y_j$ as follows
\begin{alignat}{2}
  \label{eq:Z1}
  E_i&=2\sigma z_i(\alpha)&\qquad\qquad &i=1,\dots,k\\
  E_j&=\frac{2\sigma}{\alpha}y_j(\alpha)&&j=k+1,\dots,n~.
  \label{eq:A025}
\end{alignat}
Then eqs. \eqref{eq:A009} and \eqref{eq:A011} become, respectively,
\begin{equation}
  \label{eq:Z002}
  \frac{1}{y_j}\sum_{m=0}^\infty\left\{m^{(m)}-\frac{2}{\Omega}\sum_{p=1}^{k}
    z_p^m\right\}\left(\frac{\alpha}{y_j}\right)^m+\frac{2}{\Omega}
    \sum_{\substack{p=k+1\\ p\not=j}}^n\frac{1}{y_p-y_j}+1=0
  \end{equation}
and 
\begin{equation}
  \label{eq:Z003}
  \frac{1}\Omega\sum_{\mu=1}^L\frac{\Omega_\mu}{\frac{\epsilon_\mu}{\sigma}
    -z_i}-\frac{2}{\Omega}\sum_{\substack{p=1\\ p\not=i}}^k\frac{1}{z_p-z_i}
  =\alpha+\frac{2\alpha}{\Omega}\sum_{m=0}^\infty\left(\sum_{p=k+1}^n\frac{1}{y_p^m}\right)
  (\alpha z_i)^m~.
\end{equation}
In Eqs. \eqref{eq:Z1} and \eqref{eq:A025} $z_i$ and $y_j$ are assumed to be
regular functions of $\alpha$ in some neighborhood of the origin, 
to have a finite, non-vanishing
limit when $\alpha\to0$ and to admit a Taylor expansion.

Eq. \eqref{eq:Z002} can be conveniently rewritten as
\begin{equation}
  \label{eq:B038}
  F_j\equiv\frac{1}{y_j}\sum_{m=0}^\infty \widetilde m^{(m)}[\{z_i(\alpha)\}]
  \left(\frac{\alpha}{y_j}\right)^m+\frac{2}{\Omega}
    \sum_{\substack{p=k+1\\ p\not=j}}^n\frac{1}{y_p-y_j}+1=0~,
\end{equation}
having defined
\begin{equation}
  \label{eq:B039}
  \widetilde m^{(m)}[\{z_i(\alpha)\}]= m^{(m)}-\frac{2}{\Omega}\sum_{p=1}^{k}
    z_p^m~,
\end{equation}
a form explicitly displaying the renormalization of the moments
of the level distribution induced by the dynamics of the trapped pairs.

From \eqref{eq:B038} then it clearly follows 
\begin{equation}
  \label{eq:B040}
  K_q=\sum_{j=k+1}^n y_j^q\left\{
    \frac{1}{y_j}\sum_{m=0}^\infty\widetilde m^{(m)}[z_i]
  \left(\frac{\alpha}{y_j}\right)^m+\frac{2}{\Omega}
    \sum_{\substack{p=k+1\\ p\not=j}}^n\frac{1}{y_p-y_j}+1\right\}=0~,
\end{equation}
an expression which will turn out to be useful later on.

Before examining explicitly the expansion in powers of $\alpha$
let us briefly discuss eqs. \eqref{eq:Z002} and \eqref{eq:Z003}.
We observe first of all that the case of 0 like-seniority (with no
renormalization of the moments) coincides with the findings of 
Ref.~\cite{Barbaro:2006rc} and 
is already a generalization of the case handled in ref. \cite{Yuz}
since in \eqref{eq:Z002} the moments of the 
level distribution are generic whereas in ref.~\cite{Yuz} 
the choice $\Omega_\mu=1$ is made, which is appropriate for 
a system of electrons, but not of nucleons. 
Moreover we shall show in the following that it is possible to write 
recursively (but not in a closed form!) the energy of the system associated 
with the untrapped pairs at a given order in
 $\alpha$ and eq. \eqref{eq:B038} clearly shows that at a given order $p$
only the first $p$ moments of the level distribution contribute to this energy.
At the leading order we expect
of course to recover the result of the degenerate case, the first order 
is absent because $m^{(1)}=0$ while the second is meaningful and so on.

Furthermore it turns out that the impact of the trapped pairs on the
energy of the collective state (see the eq.\eqref{eq:B038}) 
amounts to a renormalization of the moments
of the level distribution.

Likewise, for the trapped solutions at leading
order a similar effect occurs.
Indeed, at leading order (namely, $m=0$ in eq.~\eqref{eq:Z003}) 
the collective component of the state (namely the untrapped energies) 
renormalizes the coupling constant acting in the sector of the $k$ trapped
solutions according to the replacement
\begin{equation}
\alpha \rightarrow \alpha+\frac{2\alpha}{\Omega}(n-k)~.
\end{equation}

Thus we surmise the following iterative procedure: 
we first solve eq. \eqref{eq:Z002} at the leading order ($\alpha^{-1}$),
next we determine the rhs of eq. \eqref{eq:Z003} at the order $\alpha^0$
and solve the equation 
(eventually numerically), then we come back to eq. \eqref{eq:Z002} and so on.

In the next section we shall deal at leading order 
(namely in the very large coupling limit) 
with the collective component of the state energy.

\section{The collective sector}

\label{sec:2}

It is clear that when a collective state develops with a
large binding energy, then $\sigma$ has to be quite small
and, accordingly, the energy of the degenerate case,  namely
\begin{equation}
  \label{eq:A012}
  E_{\rm degenerate}=-G \left(n-\frac{v}{2}\right)
  \left(\Omega-n-\frac{v}{2}+1\right)~,
\end{equation}
should  be recovered.
We want now to show that a similar formula (i.e., up to the 
replacement $v\to v_l$) holds at the leading order in the strong coupling 
expansion for the Gaudin excited states as well.

To this purpose we go back to eq. \eqref{eq:Z002} keeping only 
the term $m=0$. Expanding $y_i(\alpha)$ as follows
\begin{equation}
  \label{eq:Y001}
  y_i(\alpha)=\sum_{h=0}^\infty\frac{1}{h!}\alpha^hy_i^{(h)}
\end{equation}
to leading order, eq. \eqref{eq:Z002} then reads
\begin{equation}
  \label{eq:A018}
  \left(1-\frac{v_l}{\Omega}\right)\frac{1}{y_j^{(0)}}
  +\frac{2}{\Omega}\sum_{\substack{m=k+1\\ m\not=j}}^n
  \frac{1}{y_m^{(0)}-y_j^{(0)}}=-1~,
\end{equation}
where only the three quantities $v_l$, $n$ and $\Omega$ (expressed by
integer numbers) appear,
while the  dependence upon the coupling constant is embedded in
the rescaling of eq. \eqref{eq:A025}.

The further rescaling 
\begin{equation}
\label{eq:A019b}
y_j^{(0)}=\left(1-\frac{v_l}{\Omega}\right) \tilde y_j
\end{equation}
leads to the system
\begin{equation}
  \label{eq:A019}  
  \frac{1}{\tilde y_j}+\frac{2}{\Omega-v_l}
  \sum_{\substack{m=k+1\\ m\not=j}}^n
  \frac{1}{\tilde y_m-\tilde y_j}+1=0~,
\end{equation}
which, redefining $\Omega$ and $n$ according to the prescriptions
\begin{eqnarray}
\Omega&\rightarrow&\Omega+v_l
\label{eq:redefO}
\\
\label{eq:redefn}
n&\rightarrow& n+\frac{v_l}{2}~,
\end{eqnarray}
can be recast as follows
\begin{equation}
  \label{eq:A020}
  f_j\equiv\frac{1}{Y_j(\Omega;n)}
  +\frac{2}{\Omega}\sum_{\substack{m=1\\ m\not=j}}^n
  \frac{1}{Y_m(\Omega;n)-Y_j(\Omega;n)}+1=0~,
\end{equation}
having set for later convenience $\tilde y_i\equiv Y_i(\Omega-v_l;n-v_l/2)$.
Note that at leading order from \eqref{eq:B038} it follows
the relation $f_j=F_j(v_l=0)$.
Eqs. \eqref{eq:A020} represent the key ingredient in describing
the dynamics of the strong coupling limit at the leading
and at the higher order as well.

To further proceed consider the equation
\begin{equation}
  \label{eq:A021}
  g_q\equiv\sum_{j=1}^n f_jY_j^q = K_q(v_l=0)=0~,
\end{equation}
whose properties are extensively described in appendix
\ref{sec:appA}. Here we first recall that, as found out in ref.\cite{Yuz},
the solutions of \eqref{eq:A021} are given by the zeros of a 
Laguerre polynomial and that the sum $\sum_{i=1}^nY_i$ can  
be analytically expressed. 
For this purpose we specify eq. \eqref{eq:A021} to the case $q=1$ and use 
\eqref{eq:A020}.  
Thus the first term of the sum yields $n$. Then the contributions 
to the sum stemming from the second term can be collected pairwise to get
\begin{equation}
\frac{2}{\Omega}\frac{Y_i}{Y_k-Y_i}+\frac{2}{\Omega}\frac{Y_k}{Y_i-Y_k}
=-\frac{2}{\Omega}
\end{equation}
and since the number of such pairs is $n(n-1)/2$ they sum up to
$-n(n-1)/\Omega$. Finally the third term yields the required quantity. 
Thus we end up with
\begin{equation}
  \label{eq:A023}
  \sum_{i=1}^nY_i=-\frac{n(\Omega-n+1)}{\Omega}~.
\end{equation}

The above sum fully determines the behaviour of {\em all} the energies of the 
collective pairs for any state of any system (i.e. with any $n$ and $v_l$)
in the strong coupling limit, yielding (we recall that $v_l=2k$)
\begin{equation}
  \label{eq:A037}
  E_i=\frac{2\sigma}{\alpha}y_j^{(0)}=
  \frac{2\sigma}{\alpha}\left(1-\frac{v_l}{\Omega}\right)
  Y_i\left(\Omega-v_l;n-\frac{v_l}{2}\right)~.
\end{equation}
The total energy of the ground state is of course the sum of the $E_i$
and owing to \eqref{eq:A023} it turns out to be
\begin{equation}
  \label{eq:13}
  \begin{split}
    E&=\frac{2\sigma}{\alpha}\left(1-\frac{v_l}{\Omega}\right)
    \sum_{i=1}^{n-v_l/2}Y_i\left(\Omega-v_l;n-\frac{v_l}{2}\right)\\
    &=-G\left(n-\frac{v_l}{2}\right)\left(\Omega-n-\frac{v_l}{2}+1\right)~,
\end{split}
\end{equation}
coinciding with \eqref{eq:A012} up to the replacement $v_l\to v$.

This result reflects the meaning of like-seniority.
We are dealing in fact with $n$ pairs all coupled to $J=0$, hence
with a zero seniority state, but the physics of the collective component of the
state is not ruled by $n$, but instead by those pairs that take part in the
setting up of the collectivity, i.e., that are not trapped. The trapped pairs turn
out to be irrelevant to the energy of the system at this
order in $\alpha$ and play essentially the same role of the broken pairs.

These are the pairs coupled to an angular momentum $J\not=0$, which set
up the seniority. They do not interact with the other ones and therefore 
are simply accounted for by
\begin{enumerate}
\item adding their unperturbed energies $2\epsilon_\mu$ to the total energy,
\item reducing each $\Omega_\mu$ by one unit each time 
  a pair coupled to $J\not=0$ lives in the $\mu^{\rm th}$ level (blocking effect)
providing the partners of the pair live on the same s.p.l.
\end{enumerate}

We conclude that the dynamics of the collective component of the states of a 
system with $n$ pairs, whatever the degree of collectivity might be, 
is ruled in the strong coupling limit by the  equations \eqref{eq:A020},
which is free of parameters, but for the integers $\Omega$, and by the scaling
law \eqref{eq:A037}.
Also worth recalling is that the Richardson's equations for 
$n$ pairs are based on the Bethe ansatz
\begin{equation}
\left| \Psi_n \right.\rangle = 
\prod_{k=1}^n \left(\sum_{\mu=1}^L \frac{C_k}{2\epsilon_\mu-E_k} 
\hat A^\dagger_\mu \right) \left| 0 \right.\rangle ~,
\end{equation}
where 
\begin{equation}
C_k = \frac{1}{ \sqrt{ \sum_{\mu=1}^L 
\frac{\Omega_\mu}{\left( 2 \epsilon_\mu - E_k \right)^2 } } }
\end{equation}
is a normalization factor and 
\begin{equation}
\hat A^\dagger_\mu = \sum_{m_\mu=-j_\mu}^{j_\mu} 
(-1)^{j_\mu-m_\mu} \hat a^\dagger_{j_\mu m_\mu} \hat a^\dagger_{j_\mu -m_\mu}
\end{equation}
the quasi-spin operator. The Bethe ansatz represents an eigenstate of the 
pairing Hamitonian if the parameters $E_k$ fulfill the RE.
As a consequence the scaling properties of the pair energies $E_k$ above 
discussed entail analogous properties for the wave function of the 
system.

\section{The trapped pairs: an example}
\label{sec:ccc}

In this Section we address the problem of computing the contribution of the
energies of the trapped pairs to the total energy of the states.
For these we have been unable to provide a strong coupling expansion,
however we show that they
fulfill a system of equations which, in leading order, decouples from the
Richardson system for the untrapped pairs.
Furthermore this system allows one to identify the unperturbed 
configuration from where each trapped contribution arises \cite{RSD}.

To see this we expand $z_i(\alpha)$ as
\begin{equation}
  \label{eq:B029}
  z_i(\alpha)=\sum_{h=0}^\infty \frac{1}{h!}z_i^{(h)}\alpha^h
\end{equation}
and using  eq. \eqref{eq:Z1}
we can rewrite eqs. \eqref{eq:Z003} at the leading order in the form
\begin{equation}
  \label{eq:A038}
  \frac{1}\Omega\sum_{\mu=1}^L\frac{\Omega_\mu}{\frac{\epsilon_\mu}{\sigma}
    -z_i^{(0)}}-\frac{2}{\Omega}\sum_{\substack{p=1\\ p\not=i}}^k
  \frac{1}{z_p^{(0)}-z_i^{(0)}}
  =0~.
\end{equation}
As above mentioned no closed form can be given for the solutions of 
(\ref{eq:A038}), however the numerical solution now only concerns the $k$ trapped pairs instead
of the full set of $n$ pairs.

We give here an example of how our approximation scheme works as compared
with the exact solution by considering a schematic model of the
lead isotope ${}^{188}Pb$.
In Table \ref{tab:1} we quote the experimental s.p.l. of 
the shell $5\hbar\omega$ and the associated energies taken from \cite{Ir-72-B}
(the zero of the energy is arbitrary).
\begin{table}[h]
  \begin{center}
    \leavevmode
    \begin{tabular}[h]{||l|l|l|l||}
      \hline\hline
      $3p_{1/2}[2]$&$\Omega_6=1$&$\epsilon_6=0$&$\epsilon_6-\bar\epsilon=
      1.897$ MeV\\
      \hline
      $2f_{5/2}[6]$&$\Omega_5=3$&$\epsilon_5=-0.57$ MeV&
      $\epsilon_5-\bar\epsilon=1.327$ MeV\\
      \hline
      $3p_{3/2}[4]$&$\Omega_4=2$&$\epsilon_4=-0.90$ MeV&
      $\epsilon_4-\bar\epsilon=0.997$ MeV\\
      \hline
      $1i_{13/2}[14]$&$\Omega_3=7$&$\epsilon_3=-1.64$ MeV&
      $\epsilon_3-\bar\epsilon=0.257$ MeV\\
      \hline
      $2f_{7/2}[8]$&$\Omega_2=4$&$\epsilon_2=-2.35$ MeV&
      $\epsilon_2-\bar\epsilon=-0.453$ MeV\\
      \hline
      $1h_{9/2}[10]$&$\Omega_1=5$&$\epsilon_1=-3.47$ MeV&
      $\epsilon_1-\bar\epsilon=-1.573$ MeV\\
      \hline\hline
    \end{tabular}
    \caption{Level structure of the highest neutron shell of lead}
    \label{tab:1}
  \end{center}
\end{table}
Observe that in the present case $\bar\epsilon=-1.897$ MeV,
a quantity to be subtracted out from the single particle energies, and
$\sigma=1.056$ MeV.

We choose, as an example,
an excited state by first switching off the interaction 
($\alpha\to\infty$) and then by filling the two lowest levels
(9 pairs) and placing two pairs in the 
level $3p_{3/2}$ and one in the $3p_{1/2}$.

In this case the system \eqref{eq:A038} contains only 3 equations.
We solved it numerically getting 
(after the shift $\epsilon_i\to\epsilon_i-\bar\epsilon$)
\begin{equation}
  \label{eq:B025}
  \begin{split}
    z_1^{(0)}&=1.719\\
    z_2^{(0)}&=0.833+0.085i\\
    z_3^{(0)}&=0.833-0.085i~,
  \end{split}
\end{equation}
in turn yielding
\begin{equation}
  \label{eq:D025}
  \begin{split}
    E_1^{(0)}&=-0.164~{\rm MeV}\\
    E_2^{(0)}&=(-2.035+0.179i)~{\rm MeV}\\
    E_3^{(0)}&=(-2.035-0.179i)~{\rm MeV}
  \end{split}
\end{equation}
for the energies $E_i$.

For later convenience we introduce also the quantity
\begin{equation}
  \label{eq:B041}
  \zeta_q(\alpha)=\sum_{i=1}^k\left[z_i(\alpha)\right]^q
\end{equation}
together with the expansion
\begin{equation}
  \label{eq:C041}
  \zeta_q(\alpha)=\sum_{h=1}^\infty \frac{\alpha^h}{h!}\zeta_q^{(h)}~.
\end{equation}
For example in the present case
\begin{equation}
  \label{eq:B042}
  \zeta_1^{(0)}=3.384~.
\end{equation}

\section{The higher order corrections}
\label{sec:x5}

In this section we use the iterative procedure previously discussed
to get the higher order corrections for $y_j(\alpha)$ and $z_j(\alpha)$.
These will be computed up to order $\alpha^4$.

We start, as in sec. \ref{sec:2},  by examining the equation $K_1=0$
(see \eqref{eq:B040}), which reads
\begin{equation}
  \label{eq:B026}
  \sum_{m=0}^\infty \widetilde m^{(m)}
  \alpha^m\sum_{i=k+1}^n\frac{1}{y_i^m}-\frac{(n-k)(n-k-1)}{\Omega}
  +\sum_{i=k+1}^ny_i=0~,
\end{equation}
having again used the same procedure used in getting
eq. \eqref{eq:A023}. Next we expand in $\alpha$. The $0^{\rm th}$ order
is already known, 
while at first order, using the expansion \eqref{eq:Y001}, we get
\begin{equation}
  \label{eq:B030}
  \sum_{i=k+1}^ny_i^{(1)}=\frac{2}{\Omega}\sum_{i=k+1}^n
  \frac{1}{y_i^{(0)}}\sum_{p=1}^kz_p^{(0)}=-\frac{2(n-k)}{\Omega-2k}
  \zeta_1^{(0)}
\end{equation}
that is sufficient for determining the correction to the total energy.
In deriving the last expression use has been made of eq. \eqref{eq:A025}
and \eqref{eq:A037} to connect the $Y_j$ with the $y_j$.
Eq.~\eqref{eq:B030} 
suggests to replace, as in \eqref{eq:redefO} and \eqref{eq:redefn},
$\Omega$ and $n$ with 
\begin{eqnarray}
\tilde\Omega&\equiv\Omega-2k\\
\tilde n&\equiv n-k .
\end{eqnarray}
In term of these natural variables eq.~\eqref{eq:B030} assumes the compact form
\begin{equation}
  \label{eq:D030}
  \sum_{i=k+1}^ny_i^{(1)}=-\frac{2\tilde n}{\tilde\Omega}
  \zeta_1^{(0)}~.
\end{equation}

Thus the leading order correction to the state collective energy
solely arises from the presence of the trapped pairs that merely 
renormalize $\bar\epsilon$.

For the higher order terms we need to solve the equations from $K_{-l+1}=0$ to $K_1=0$:
these fix the $l^{\rm th}$ order.
To lighten the notations we introduce the coefficient
\begin{equation}
  \label{eq:D001}
  C_p=\left(\frac{\Omega}{\Omega-2k}\right)^{p+1}\frac{(n-k)(\Omega-n-k)}
  {\prod_{m=1}^p (\Omega-2k-m)}=\left(\frac{\Omega}{\tilde\Omega}\right)^{p+1}
  \frac{\tilde n(\tilde\Omega-\tilde n)}
  {\prod_{m=1}^p (\tilde\Omega-m)}~.
\end{equation}

We thus find
\begin{eqnarray}
  \label{eq:W001}
  \sum_{i=k+1}^ny_i^{(2)}&=&-2C_1
  \left\{m^{(2)}-\frac{2}{\Omega}\zeta_2^{(0)}
    -\frac{4}{\Omega\tilde\Omega}(\zeta_1^{(0)})^2
    \right\}-\frac{4\tilde n\zeta_1^{(1)}}{\tilde\Omega}\\
  \sum_{i=k+1}^ny_i^{(3)}&=&6C_2
  (\Omega-2n)\Biggl\{m^{(3)}-\frac{2}{\Omega}\zeta_3^{(0)}
  +\frac{6}{\tilde\Omega}\zeta_1^{(0)}
  \left(m^{(2)}-\frac{2}{\Omega}\zeta_2^{(0)}\right)\\
  \nonumber
  &&-\frac{16}{{\Omega\tilde\Omega}^2}\left[\zeta_1^{(0)}\right]^3\biggr\}
  +12C_1\frac{1}{\Omega}\left(\zeta_2^{(1)}+\frac{4}{\tilde\Omega}
    \zeta_1^{(0)}\zeta_1^{(1)}\right)
  -6\frac{\tilde n}{\tilde\Omega}\zeta_1^{(2)}\\
  \sum_{i=k+1}^ny_i^{(4)}&=&-24C_3\left\{{\tilde\Omega}^2-\frac{\tilde n
      (\tilde\Omega-\tilde n)(5\tilde\Omega-6)}
    {\tilde\Omega-1}\right\}\times\\
  \nonumber
  &&
  \times\left\{m^{(4)}-\frac{2}{\Omega}\zeta_4^{(0)}+\frac{8}{\tilde\Omega}
    \left[m^{(3)}-\frac{2}{\Omega}\zeta_3^{(0)}\right]\zeta_1^{(0)}\right\}\\
  \nonumber
  &-&192C_3\Biggl\{5+\frac{5\tilde n(\tilde n-6)}{\tilde\Omega}
  +30\frac{\tilde n^2}{\Omega^2}-\frac{(5\tilde n-1)(\tilde n-1)}
  {\tilde\Omega-1}\nonumber\\
  &&
  -\frac{\tilde n(\tilde n-1)}{(\tilde\Omega-1)^2}
  \Biggr\}\left\{\left[m^{(2)}-\frac{2}{\Omega}\zeta_2^{(0)}\right]
    \left[\zeta_1^{(0)}\right]^2
    -\frac{2}{\Omega\tilde\Omega}
    \left[\zeta_1^{(0)}\right]^4\right\}
  \nonumber\\
  \nonumber
  &+&144C_3\frac{(\tilde\Omega-3)(\Omega-2n)}{\Omega}\left[
    m^{(2)}-\frac{2}{\Omega}\zeta_2^{(0)}\right]\zeta_1^{(1)}\\
  \nonumber
  &+&24C_3\tilde\Omega\Biggl\{2\tilde\Omega-9\tilde n-1+
    12\frac{\tilde n}{\tilde\Omega}-\frac{(3\tilde n-1)(\tilde n-1)}
    {\tilde\Omega-1}
  \\
  \nonumber
  &&-\frac{\tilde n(\tilde n-1)}{(\tilde\Omega-1)^2}\Biggr\}
  \left[m^{(2)}-\frac{2}{\Omega}\zeta_2^{(0)}\right]^2\\
  \nonumber
  &-&48C_2\frac{\Omega-2n}{\Omega}\left\{\zeta^{(1)}_3+\frac{6}{\tilde\Omega}
  \zeta_1^{(0)}\zeta_2^{(1)}+\frac{24}{\tilde\Omega^2}\zeta_1^{(1)}
  \left[\zeta^{(0)}_1\right]^2
  \right\}\\
  \nonumber
  &+&24C_1\frac{1}{\Omega}\left\{\zeta_2^{(2)}+\frac{4}{\tilde\Omega}
    \zeta_1^{(0)}\zeta_1^{(2)}+\frac{4}{\tilde\Omega}
      \left[\zeta_1^{(1)}\right]^2\right\}
    -\frac{8\tilde n}{\tilde\Omega}\zeta_1^{(3)}~.
\end{eqnarray}

It is worth to point out the drastic simplification occurring 
for the unique state having all the pairs untrapped (vanishing like-seniority). 
Indeed in this case:
\begin{eqnarray}
  \label{eq:W002}
  \sum_{i=k+1}^ny_i^{(1)}&\to&0\\
  \sum_{i=k+1}^ny_i^{(2)}&\to&-2
  \frac{n(\Omega-n)}{\Omega-1}m^{(2)}\\
  \sum_{i=k+1}^ny_i^{(3)}&\to&6\frac{n(\Omega-n)(\Omega-2n)}
  {(\Omega-1)(\Omega-2)}m^{(3)}\\
  \sum_{i=k+1}^ny_i^{(4)}&\to&24\frac{n(\Omega-n)}{
    (\Omega-2)(\Omega-3)}\Biggl\{\left[\frac{n(\Omega-n)(5\Omega-6)}{(\Omega-1)^2}
  -\frac{\Omega^2}{\Omega-1}\right]m^{(4)}\\
  \nonumber
  &&
  +\left[2\Omega+1+\frac{n(4-9(\Omega-n))}{\Omega-1}
    -\frac{(4n-1)(n-1)}{(\Omega-1)^2}\right.\\
  \nonumber
  &&
  \left.-\frac{n(n-1)}{(\Omega-1)^3}\right]\left[m^{(2)}\right]^2\Biggr\}~,
\end{eqnarray}
which coincides with the findings of ref.~\cite{Barbaro:2006rc}.

Two comments are now in order. First in the above we have explicitly 
inserted the second moment of the level distribution, although its
value is 1 by definition, in order to explicitly follow how the moments
of the s.p.l. distribution are renormalized order by order.  
Next we recall that $m^{(3)}$ coincides with the 
skewness of the distribution and $m^{(4)}$ is linked to the kurtosis $c$
by the relation $c+3=m^{(4)}$.
We thus see that the coefficients of the strong coupling expansion of the ground 
state energy reflect finer and finer details of the levels distribution as the order 
grows. The same occurs for the excited states, but here the connection is much more 
cumbersome.

Now we switch to the trapped states and we evaluate numerically,
order by order, the
unknown quantities $z_i^{(m)}$. We have already determined, 
in the previous section, the zero order $z_i^{(0)}$ by solving numerically 
eq. \eqref{eq:A038}. At the next-to-leading order eq. \eqref{eq:Z003} reads
\begin{equation}
  \label{eq:B031}
  \begin{split}
  \frac{1}{\Omega}\sum_{\mu=1}^L\frac{\Omega_\mu}{\left(
      \frac{\epsilon_\mu}{\sigma}-z^{(0)}\right)^2}
  +\sum_{\substack{p=1\\ p\not=i}}^k\frac{z^{(1)}_p-z^{(1)}_i}{
    (z^{(0)}_p-z^{(0)}_i)^2}&=1+\frac{2}{\Omega}\sum_{j=k+1}{n}\frac{
    1}{y^{(0)}_j}\\
  &=1+k-n~,
  \end{split}
\end{equation}
where in the second line use has been made of eq. \eqref{eq:K001} 
of Appendix B.
The above is now a linear sistem. 
We solved it within our model, getting the results collected 
in Table \ref{tab:2}, where also the terms up to the fourth order are reported.

In accord with the previous discussion, the 0-th order contribution
to the $z_i$ relates to $G=\infty$ ($\alpha=0$) and the higher order terms describe the evolution with $G$ of the trapped energies.
For these the impact of the untrapped pairs is felt.

Concerning the range of validity of the expansion (\ref{eq:B029}), it should
be set by the critical values of $G$ (or $\alpha$) which are specific of
each state of the pairing Hamiltonian.

\begin{table}[h]
  \begin{center}
    \leavevmode
    \begin{tabular}[h]{||c||r|r|r||}
      \hline\hline
      &$i=1$&$i=2$&$i=3$\\
      \hline
      $z^{(1)}_i$&$-0.015$&$-0.025+0.012i$&$-0.025-0.012i$ \\
      \hline
      $z^{(2)}_i/2!$&$0.181$&$0.178+0.081i$&$0.178-0.081i$\\
      \hline
      $z^{(3)}_i/3!$&$0.124$&$0.076+0.013i$&$0.076-0.013i$\\
      \hline
      $z^{(4)}_i/4!$&$-0.747$&$-0.251+0.034i$&$-0.251-0.034i$\\
      \hline\hline
    \end{tabular}
    \caption{The results of the numerical calculation for the 
      higher order coefficients $z^{(p)}_i$ in the expansion of $z_i$. 
      The index $p$ refers to the order of the correction. The index $i$ runs over the
      three trapped pairs living in the $3p_{1/2}$ ($i=1$) and $3p_{3/2}$ ($i=2,3$) 
      s.p.l.}
    \label{tab:2}
  \end{center}
\end{table}

\section{Comparison with the exact results}
\label{sec:8}

In this Section we test the efficiency of the formalism previously developed 
by comparing its predictions with the exact results obtained by numerically
solving the RE in the specific example of our toy model for $^{188}Pb$.

We also search for the range of values of the coupling constant $\alpha$ 
(or $G$) where our strong coupling expansion holds valid.
This we do by discussing the analytic properties in $G$ of the solutions $E_i$ 
($i=,1\cdots n$) (and hence of the system's total energy $E$) of the RE.

\subsection{The singularities in $G$ of the pairs and of the total energies}
\label{sec:sing}

\begin{figure}[h]
  \begin{center}
    \leavevmode
    \hbox{
 \hspace{-1.5cm}
      \epsfig{file=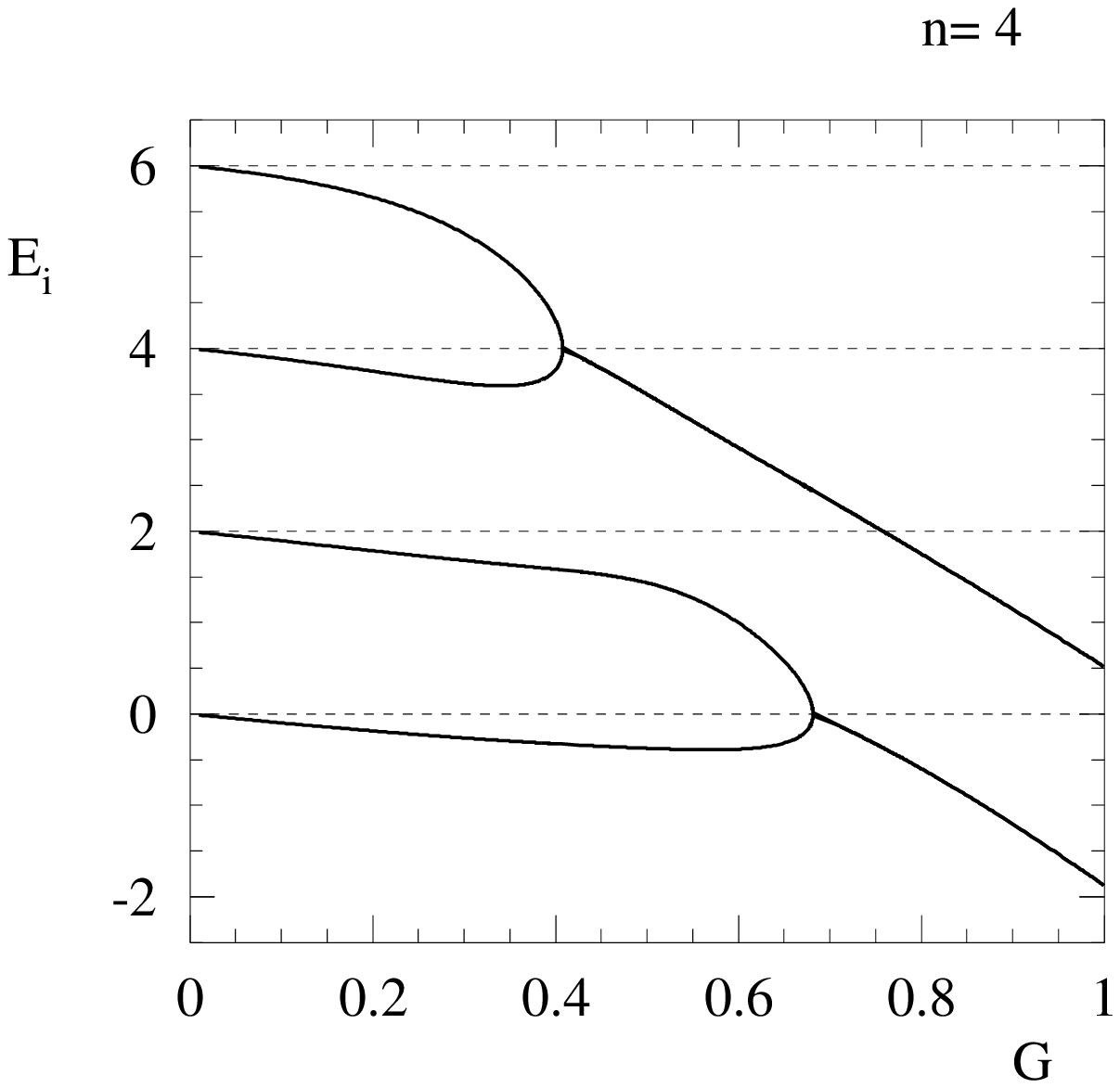,width=9.cm,height=10cm}
 \hspace{-2.cm}
      \epsfig{file=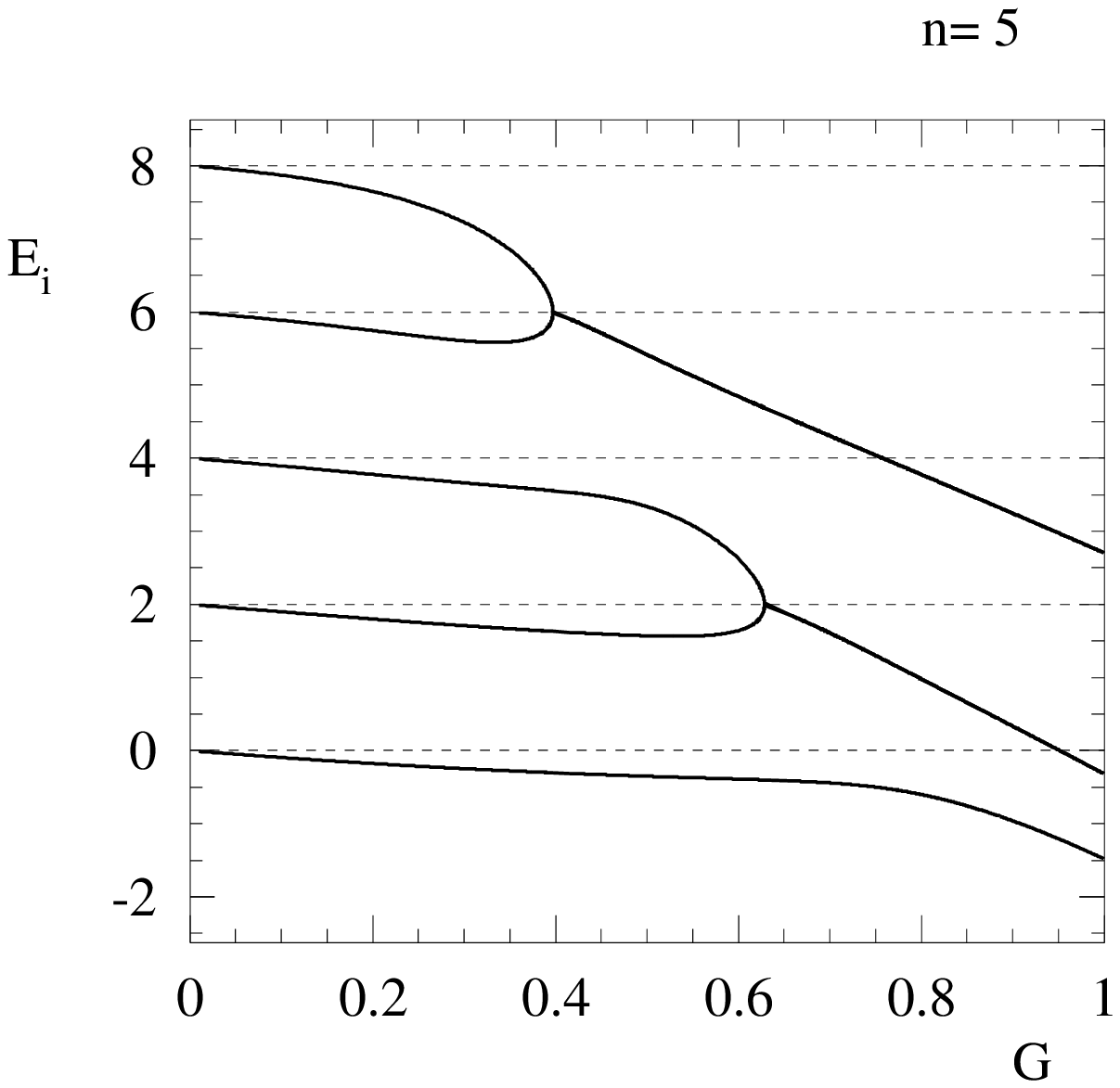,width=9.cm,height=10cm}
      }
    \caption{Evolution with $G$ of the real parts of the single particle energies for a system 
        with equally spaced unperturbed levels with unit pair degeneracy, with 12 levels and with 
        4 pairs (left panel) and 5 pairs (right panel). $G$ and $E_i$ in arbitrary units. }
  \label{fig:3}
    \end{center}
  \end{figure}
It was discovered by Richardson \cite{Ri-65} that some critical values of $G$ 
may exist where a level with pair degeneracy $\Omega_\mu$
tries to host $\Omega_\mu+1$ pairs. 
These critical points necessarily appear in the case of the metals. 
In fact here for $G\to 0$ the pair energies $E_i$ tend
to the unperturbed values $2 \epsilon_i$, which are real.
But to reach the collective state at high $G$ all the $E_i$, 
but the lowest one,
must escape from the grid set up by the poles displayed by the RE, which are 
placed at the unperturbed single particle energies.
As it is well-known, the escaping mechanism relates to the evolution with $G$
of the pair energies.
Considering a specific $E_i$, note that 
it starts from the real value $2\epsilon_i$ at $G=0$ and then
merges with the lower neighbour solution $E_{i-1}$ at the energy 
$2\epsilon_{i-1}$ for a particular critical value of $G$. 
Beyond this critical point the two pair energies $E_i$ and $E_{i-1}$ become 
complex conjugate and their imaginary part enable them to overcome
all the other obstacles to their evasion from the grid. 
This mechanism is shown in figs. \ref{fig:3}
for a typical case of equally spaced unperturbed levels with unit pair 
degeneracy.

One would expect these critical values of $G$ to play a crucial role in
determining the convergence domain of the strong coupling expansion
for the system's energy.
Actually the situation turns out to be more involved since the singularities 
of the pair energies $E_i$ cancel out in the sum yielding the total energy of 
the system (see Ref.~\cite{Barbaro:2006rc}
 for a discussion of this point).

Clearly the situations 
occurring in nuclear physics are drastically different from the metallic
situation since the s.p.l. energies are different
and must be examined case by case. 

\begin{figure}[htb]
  \begin{center}
    \leavevmode
    \epsfig{file=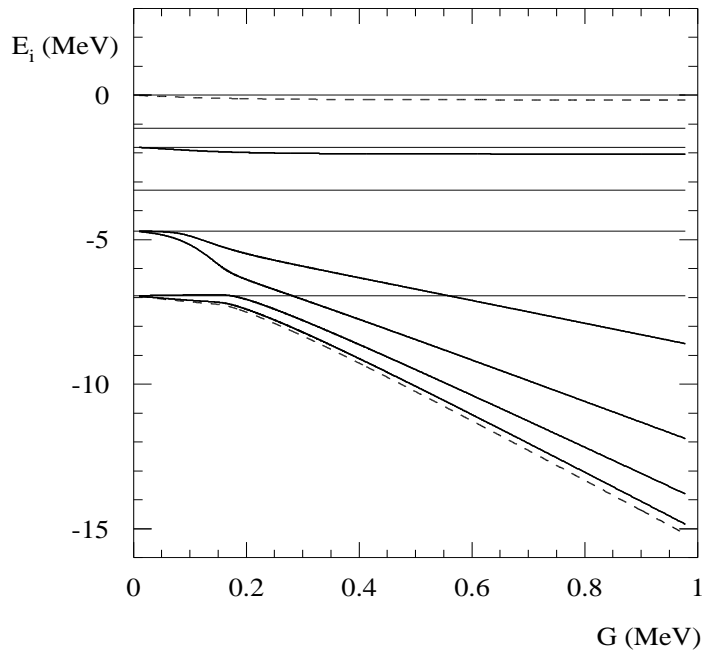,width=12cm,height=11.cm}
    \caption{The real part of the solutions $E_i$ for the
      fifth shell of ${}^{188}Pb$. Solid lines represent the real part of
      two complex conjugate solutions, dashed lines represent single real 
      solutions.}
    \label{fig:1}
  \end{center} 
\end{figure}

Sticking to our example of $^{188}Pb$ we have drawn in fig. \ref{fig:1} 
the real parts of the exact solutions of the RE.

In this connection we remind that our model of $^{188}Pb$ has 12 pairs
in the $5\hbar\omega$ shell.
For sake of illustration we consider of this nucleus the excited state
with $N_G=3$ (or $v_l=6$). Hence 3 pairs remain trapped: they arise 
from the $3p_{3/2}$ and $3p_{1/2}$ levels.
Of the remaining 9, 5 pairs arise from the $1h_{9/2}$ level and 4 from the
$2f_{7/2}$ one.
It is found that at very small $G$ the energies of the former are proportional 
to the fifth roots of the unity. 
Thus 4 of them are complex and the $5^{\rm th}$ is real.
Since all of them decrease as $G$ increases, they reach their asymptotic 
values at high $G$ without encountering any singular point.

Concerning the energies of the pairs stemming from the $2f_{7/2}$ level, 
they are two by two complex conjugate and hence not affected by the trapping 
mechanism.
Thus no singularities in $G$ arise and one would expect the expansion 
in $\alpha$ for the collective part of the energy to converge everywhere.
Note however that this analysis refers to $G$ positive and real.

\begin{figure}[htb]
  \begin{center}
    \leavevmode
    \epsfig{file=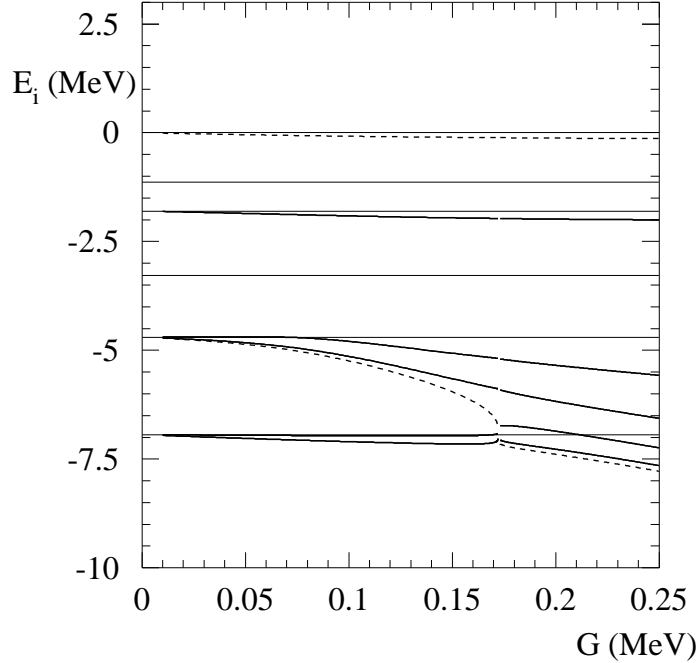,width=11cm,height=11cm}
    \caption{The real part of the solutions $E_i$ for the
      fifth shell of ${}^{188}Pb$ with the sub-shells $2f_{7/2}$ and $1h_{9/2}$
        interchanged. Solid and dashed lines as in fig. \protect\ref{fig:1}.}
    \label{fig:2}
  \end{center}
\end{figure}

To better illustrate the subtelties of the escaping mechanism
we next consider another case (not realistic)
by interchanging the levels $2f_{7/2}$ and $1h_{9/2}$, as shown in fig. 
\ref{fig:2}.
Now  the lowest level may host only 4 pairs that could reach their asymptotic
value with continuity, but the next one has 5 pairs and one of the associated
$E_i$ has to be real. Thus it cannot escape the trapping, unless through
a critical point, that must necessarily exist since the solutions
of the RE in the strong coupling regime has 9 pairs taking part
to the collectivity. Thus at this critical 
value this single solution must meet the four
lower ones exactly at the lowest unperturbed level
as fig. \ref{fig:2} indeed shows to happen (note that for $G>G_{\rm cr}$
the four lowest solutions are accordingly complex and  hence only
two lines appear in our figure, hardly distinguishable, however, because
they are very close to each other).
One could be tempted to conclude that the validity of the power expansion 
in $\alpha$ ends 
at $\alpha=\frac{2\sigma}{G_{\rm cr}\Omega}$, but, as we shall see in the
next Section, this is not so.

Actually precise statements
about the domain of convergence of the strong coupling series are hard to 
make (see, however, Ref.~\cite{Barbaro:2006rc}) 
and in fact each case should be separately examined.
Concerning the existence of critical values of $G$, they depend crucially upon the 
occupation number of the levels.

\begin{figure}[htb]
  \begin{center}
    \leavevmode
    \epsfig{file=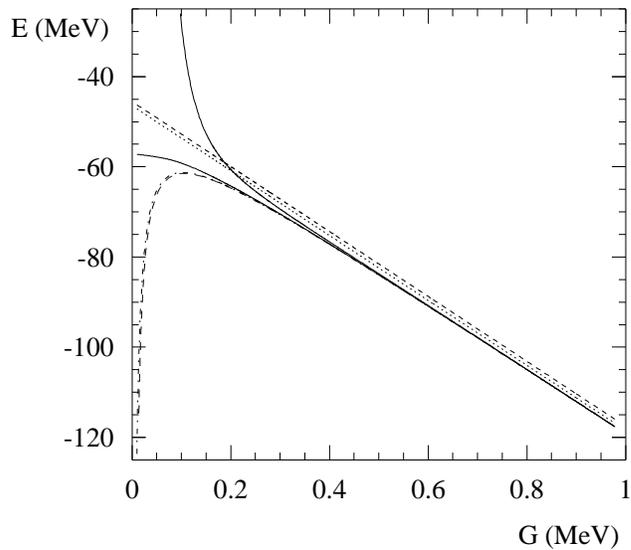,width=10.5cm,height=10.5cm}
    \caption{The exact solution for the pairing energy of ${}^{188}Pb$
      (solid line), compared with the order $G$ (dashed line), the order
      $G^0$ (dotted line), the order $G^{-1}$ (dash-dotted line), the order
      $G^{-2}$ (long-dashed line) and the order $G^{-3}$ (solid line again).
      The last three contributions are divergent at the origin.}
    \label{fig:4}
  \end{center}
\end{figure}

\subsection{Testing our approach}

In this subsection we test our approach against the exact solution of the RE.
We display in fig. \ref{fig:4} the exact result for the case of the
excited state of $^{188}Pb$
and compare it with the expansion in powers
of $\alpha$ up to the order $\alpha^3$ (or $G^{-3}$).

In the figure we have accounted for an overall energy shift since in our model 
$\bar\epsilon\not=0$. First we observe that the order next to the leading 
is not vanishing owing to the interaction between the collective mode and 
the three trapped pairs, but the effect appears to be very small 
(indeed the two lines representing the $0^{th}$ and $1^{st}$ order 
are almost superimposed).
Next it is seen that a very good accord between the RE exact 
solution and our approach is obtained up to $G\cong0.35$: for lower
values of $G$ it appears that higher order terms in the expansion are required.

However, and importantly, for $G\simeq 0.3$ the terms of the expansion diverge,
thus possibly signalling the occurrence of a singularity 
(see Ref.~\cite{Barbaro:2006rc}).
Note that the highest order in the expansion is the most sensitive to the occurrence
of this possible singularity.

This outcome might be related to the well-known result for the energy of
a pair living in two levels, a case where obviously critical values of
$G$ cannot exist, which reads
\begin{equation}
E=-d(\lambda+\sqrt{1+\lambda^2})=-G-\sqrt{d^2+G^2}~,
\label{eq:1p2l}
\end{equation}
being $G=\lambda d$ and $d$ the distance between the two levels.
Clearly \eqref{eq:1p2l} can be expanded only for $\lambda > 1$.
Since in our model the average 
$d$ is also approximately 1, one may conclude that for the excited state of
$^{188}Pb$ the strong
coupling expansion holds valid till values of $G$ much smaller than in the
case of one pair living in two levels (assuming the same $G$ and $d$ in both
situations).
Of course it should be kept in mind that we deal with an {\em excited} 
(and not the {\em ground}) state of $^{188}Pb$ where we have six (and not one)
pairs contributing to the collective part of the 
energy and where the pair degeneracy of the s.p.l.
is not one.

In fact in general for the ground state 
(see Ref.~\cite{Barbaro:2006rc}) it turns out that
the larger the pair number $n$ is,
the larger the domain of validity of the strong coupling expansion becomes:
in the BCS case, e.g., which corresponds to the thermodynamic limit,
the expansion holds for 
$\lambda>1/\pi$ (or $G>d/\pi$).
Note however that in BCS an infinite number of $G_{\rm crit}$ exists filling 
the range $0\leq G_{\rm crit} \leq 1.13 d$~\cite{RSD1}.

In conclusion while we cannot make a precise statement about the radius of
convergence of the strong coupling series (each case requires to be
separately examined) our results confirm that a singularity
exists in the complex $G$-plane, limiting the validity of the strong coupling 
expansion to values of 
$G>{G}^{\rm sing}$, being $G^{\rm sing}$ the smallest element of the set of all the
singular points. Notably, $G^{\rm sing}$ is 
unrelated to the $G_{\rm crit}$, in
fact existing even when there are no $G_{\rm crit}$, and, furthermore,
that our value of $G^{\rm sing}$ appears to be in accord with the finding of
Ref.~\cite{Barbaro:2006rc}.

\section{Conclusions}
\label{sec:6}

In this paper the RE, which solve the pairing Hamiltonian 
problem for a system of $n$ pairs, 
are reduced, in the strong coupling limit, to a parameter-free
set of equations (but for the total pair degeneracy of the s.p.l.,
$\Omega$),
namely the set \eqref{eq:A020}. 

The eigenvalues of this system are obtained by solving
an algebraic equation of order $n$, whose coefficients are explicitly
given in eq. \eqref{eq:A029} of Appendix A.

Once the solutions of the system are known, the physical
unknown $E_j$ are determined to leading order up to a rescaling. 
In the case of states of 0 like-seniority ($N_G=0$) this rescaling is indeed 
what one would naturally expect, but the rescaling is quite more involved
for states of finite $v_l$.
Finally, and remarkably, the total energy of the state can be determined 
to leading order without solving explicitly the system.

Addressing the higher order corrections, we have explicitly derived
their analytic expressions in the simpler case $v_l=0$.
We have shown, in accord with ref.~\cite{Barbaro:2006rc},
that at the order $p$ only $p$ parameters
are involved in their determination, 
namely the first $p$ moments of the s.p.l. distribution.
When $v_l\neq 0$ the trapped states renormalize dynamically these moments.

Concerning the numerical aspect of our approach,
in the general case only one substantial calculation is 
required, the remaining steps to get the energies involving the solution 
of simple linear systems.
Thus we point out that, although an exact analytic expression for
the expansion of the whole energy of the excited states cannot be provided 
because of the coupling
of the trapped and untrapped solutions, however order by order in our expansion
the two systems of equations yielding the energies of the trapped and untrapped pairs
can be decoupled.
As a consequence on the one side the collective part of the energy of
{\em any} state can indeed be expressed as an expansion and on the other
this occurrence offers numerical advantages, especially when $n$ is large,
and also a better insight on the nature of the excited states of the
pairing Hamiltonian.

Finally the convergence of the series is discussed. 
Its radius of convergence is set by a singularity lying in the complex
plane of the coupling constant, whose exact location crucially depends
upon the distribution of the levels. However the modulus of this singularity
must be smaller than the lowest $G_{\rm crit}$.
Its physical meaning represents the minimum value of $G$ at which the single 
particle aspect of the problem can be treated as a perturbation.

\appendix

\section{Properties of the functions $Y_i(\Omega;n)$}
\label{sec:appA}

We study the properties of the system \eqref{eq:A021}:
$$\sum_{i=1}^n\left[Y_i(\Omega;n)\right]^{n-1}
+\frac{2}{\Omega}\sum_{i=1}^n
\sum_{\substack{k=1\\ k\not=i}}^n
\frac{\left[Y_i(\Omega;n)\right]^{n}}{Y_k(\Omega;n)-Y_i(\Omega;n)}
+\sum_{i=1}^n\left[Y_i(\Omega;n)\right]^{n}=0~.$$

\begin{description}
\item[Property I:] The system \eqref{eq:A021} is algebraic.
  
  In fact for each term of the second sum, let it be $\frac{2}{\Omega}
  \frac{Y_i^m}{Y_k-Y_i}$ another term in the sum exists with the indices
  interchanged, namely $\frac{2}{\Omega}
  \frac{Y_k^m}{Y_i-Y_k}$ and their sum is clearly a polynomial. The whole 
  second term in \eqref{eq:A021} is thus a symmetric function 
  of order $m-1$ of the variables $Y_i$.
\item[Property II:] each equation of the system \eqref{eq:A021}
  can be expressed in terms of the symmetric polynomials
  \begin{equation}
    \label{eq:A022}
    S_k(\Omega,n)=\sum_{p_1<p_2<\dots<p_k}Y_{p_1}Y_{p_2}\dots Y_{p_k}~.
  \end{equation}  
\item[Property III:]
\begin{equation}
  \sum_{i=1}^nY_i=-\frac{n(\Omega-n+1)}{\Omega}~.
\end{equation}
\item[Property IV:] if $n=\Omega-1$ the system \eqref{eq:A021}
  admits the solution $Y_i=0~\forall i$. In fact only the equation $g_1$
  contain a constant term that reads (see eq. \eqref{eq:A023})
  $-n(\Omega-n+1)/\Omega$.
\item[Property V:] if $k$ solution coincide they vanish and further 
  must be
  $\Omega=k-1$; conversely if $\Omega=k-1$ there exist $k$ vanishing solutions.
  
  To prove this we rewrite the system \eqref{eq:A020} labelling from 1 to
  $k$ the vanishing $Y_i$ and we
  separate the system $f_i=0$ into two subsystems.
  The first  reads
  \begin{alignat}{3}
    \label{eq:A026}
    \frac{1}{Y_i}+\frac{2}{\Omega}\sum_{\substack{m=1\\ m\not=i}}^k
    \frac{1}{Y_m-Y_i}+\frac{2}{\Omega}
    \sum_{m=k+1}^n\frac{1}{Y_m}&=-1~,&~~~~~~~&i=1,\dots,k
    \intertext{and the second is}
    \label{eq:A027}
    \frac{1-\dfrac{2k}{\Omega}}{Y_j}+\frac{2}{\Omega}\sum_{\substack{p=k+1\\ 
        p\not=j}}^n
    \frac{1}{Y_m-Y_j}&=-1~,&~~~~~~~&j=k+1,\dots,n~.
  \end{alignat}
  Imagine now we solve in some way, numerically for instance, the set
  \eqref{eq:A027}. The quantity $\sum_{m=k+1}^n\frac{1}{Y_m}$ in
  \eqref{eq:A026} will thus be a finite, known term.
  Rescaling then the $Y_i$ in \eqref{eq:A026} according to
  \begin{equation}
    \label{eq:A028}
    Y_i=\dfrac{\tilde Y_i}{1+\dfrac{2}{\Omega}\sum_{m=k+1}^n\dfrac{1}{Y_m}}
  \end{equation}
  the equations for the $\tilde Y_i$ will now keep exactly the form 
\eqref{eq:A020},
  but we shall have to deal with only $k$ of them. Thus repeating
  the  derivation of property V we conclude
  that these (and consequently the $Y_i$)
  can vanish only if
  $$\Omega=k-1~.$$
\item[Property VI:] the solutions $Y_i$ of the system \eqref{eq:A021}
  are the roots of the equation (in $x$)
  \begin{equation}
    \label{eq:A029}
    \sum_{p=0}^n \binom{n}{p}\frac{(\Omega-n+1)_p}{\Omega^p}x^{n-p}=0~.
  \end{equation}

  In fact we know that solving the system \eqref{eq:A021} amounts to find
  the roots of the algebraic equation
  \begin{equation}
    \label{eq:A030}
    \sum_{p=0}^n (-1)^p S_p x^{n-p}=0
  \end{equation}
  with the $S_p$ defined by \eqref{eq:A022} and $S_0=1$. Owing to Property II
  each $g_m$ can be expressed in terms of the $S_p$ with $p=1,\dots,m$
  and $S_m$ is contained linearly.
  Thus $g_1$ can only contain linearly $S_1$, that is immediately determined,
  and all the other $S_k$ can be obtained recursively by solving first order 
  equations, thus getting the $S_p$ as functions of $\Omega$ and $n$.
  
  Further, it is easily seen by induction that
  \begin{equation}\label{eq:A031}
    S_p=\frac{P_p(\Omega)}{\Omega^p}
  \end{equation}
  where $P_p$ is a 
  polynomial (to be determined later) of order $p$ in $\Omega$.

  To further determine $P_p(\Omega)$,
  property VI tells us that if $\Omega=k-1$ then $k$ solutions are vanishing
  and thus the first $k$ coefficients of the 
  equation \eqref{eq:A030} must vanish.

  For instance for $\Omega=n-1$
  the equation must have the form $x^n=0$, entailing the vanishing of
  all the polymial. 
  Thus all of them must factorize a term $(\Omega-n+1)$.
  This fixes $P_1$ up to a constant. At $\Omega=n-2$ all the polynomial
  but the first must vanish in order to have
  $n-1$ vanishing solutions, and so on. Thus they  ultimately
  take the form
  \begin{equation}
    \label{eq:A032}
    P_k(\Omega)=t_k^n (\Omega-n+1)(\Omega-n+2)\dots(\Omega-n+k)=(\Omega-k+1)_k
  \end{equation}
  (having introduced the Pochhammer symbol).
  In the above the $t_k^n$ are numerical (rational) 
  coefficients that can depend 
  upon $n$, but {\em not} upon $\Omega$.

  To fix them we exploit their independence from $\Omega$ and take
  the limit $\Omega\to\infty$. Then from \eqref{eq:A031} and \eqref{eq:A032}
  it follows
  \begin{equation}
    \label{eq:A033}
    S_k(\Omega\to\infty,n)=t^n_k~.
  \end{equation}
  On the other hand in this limit the system \eqref{eq:A020} 
  is trivially solved,
  because the Pauli terms vanish, and yields 
  $Y_i(\Omega\to\infty,n)=-1$. Thus in this limit eq. \eqref{eq:A030}
  becomes
  \begin{equation}
    \label{eq:A034}
    \sum_{p=0}^n (-1)^p t_p^n x^{n-p}=(x+1)^n=\sum_{p=0}^n
    \binom{n}{p}x^{n-k}=0~,
  \end{equation}
  that immediately provides
  \begin{equation}
    \label{eq:34}
    t^n_p=(-1)^p\binom{n}{p}~.
  \end{equation}
  Thus \eqref{eq:A030} can be rewritten in compact form as
  \begin{equation*}
    \sum_{p=0}^n \binom{n}{p}\frac{(\Omega-n+1)_p}{\Omega^p}x^{n-p}=0~.
  \end{equation*}
  This completes the proof of the property.
\item[Property VII:] the functions $Y_i(\Omega;n)$ display branch points
  for the integers $\Omega=1,2,\dots,n-1$ and eventually a pole for $\Omega=0$.
  Near $\Omega=k-1$ they behaves like $Y_i(\Omega;k-1)\propto\sqrt[k]{\Omega}$.
  If we put $\Omega=k-1+\epsilon^k$ with $k\to0$ we find that the
  solutions $Y_i$ have the behaviour
  \begin{equation}
    \label{eq:A035}
    Y_i(k-1+\epsilon^k;n)=\sum_{p=1}^k 
    C_p^{(k)}\left[e_i^{(k)}\epsilon\right]^p
  \end{equation}
  where the $e_i^{(k)}$ are the roots of the unity, namely
  \begin{equation}
    \label{eq:A036}
    e_m^{(k)}=e^{2i\pi\frac{m}{k}}~.
  \end{equation}
  
\end{description}

\section{Useful relations}

We display here a list of properties of the functions $y_i(\alpha=0)$ 
derived from
the equations 
$$g_m=0$$
(with $m$, if it is the case, $<0$). We recall that the functions $y_i$ --
with like-seniority (Gaudin number) $\not=0$ --
are obtained from
$$y_i=\left(1-\frac{2k}{\Omega}\right)Y_i(\Omega-2k;n-k)~.$$
They read
\begin{eqnarray}
  \label{eq:K001}
  \sum_{i=1}^n\frac{1}{Y_i(\Omega;n)}&=&-n\\
  \sum_{i=1}^n\frac{1}{Y_i^2(\Omega;n)}&=&
  \frac{n(\Omega-n)}{\Omega-1}\\
  \sum_{i=1}^n\frac{1}{Y_i^3(\Omega;n)}&=&
  \frac{n(\Omega-n)(\Omega-2n)}{(\Omega-1)(\Omega-2)}\\
  \sum_{i=1}^n\frac{1}{Y_i^4(\Omega;n)}&=&
  \frac{n(\Omega-n)}{(\Omega-1)(\Omega-2)(\Omega-3)}
  \left[\Omega^2-\frac{n(\Omega-n)(6\Omega-5n)}{\Omega-1}\right]\\
  \sum_{i=1}^n Y_i(\Omega;n)&=&-\frac{n(\Omega-n+1)}{\Omega}\\
  \sum_{i=1}^n Y_i^2(\Omega;n)&=&-\frac{n(\Omega-2n+2)(\Omega-n+1)}{\Omega^2}\\
  \sum_{i=1}^n Y_i^3(\Omega;n)&=&-\frac{1}{\Omega^3}n(\Omega-n+1)\\
  &&\times\left[6-11n+5n^2+\Omega(5-5n+\Omega)\right]~.
  \nonumber
\end{eqnarray}


\end{document}